\documentclass[aps,prd,twocolumn,showpacs,floatfix,nofootinbib,superscriptaddress]{revtex4-1}
\usepackage{dcolumn}

\usepackage{amsmath, amsfonts, amssymb, mathrsfs, times, float, color, bm}
\usepackage{extarrows}
\usepackage{graphicx}
\usepackage{float}
\usepackage{graphicx,epstopdf}
\usepackage{dcolumn}
\usepackage{bm}
\usepackage{exscale}
\usepackage{relsize}
\usepackage[colorlinks=true,breaklinks=true,linkcolor=blue,citecolor=blue,urlcolor=blue]{hyperref}
\usepackage{soul}
\newcommand{\nn}{\nonumber}

\begin{document}

\title{Gravitational deflection of relativistic massive particles by wormholes}
\author{Zonghai Li}
\affiliation{School of Physical Science and Technology, Southwest Jiaotong University, Chengdu, 610031, China}
\author{Guansheng He}
\affiliation{School of Mathematics and Physics, University of South China, Hengyang, 421001, China}
\author{Tao Zhou}
\email{taozhou@swjtu.edu.cn}
\affiliation{School of Physical Science and Technology, Southwest Jiaotong University, Chengdu, 610031, China}

\date{\today}

\begin{abstract}
In this paper, the gravitational deflection of relativistic massive particles up to the second post-Minkowskian order by static and spherically symmetric wormholes is investigated in the weak-field limit. These wormholes include the Janis-Newman-Winicour wormhole, a class of zero Ricci scalar scalar-tensor wormholes, and a class of charged Einstein-Maxwell-dilaton wormholes. With the Jacobi metric approach, the Gauss-Bonnet theorem is employed to study the gravitational deflection. In this scheme, the deflection angle as a topological effect is considered. Moreover, we analyze the influence of the spacetime parameters on the results.
\end{abstract}


\maketitle

\section{Introduction}
Gravitational lensing is one of the most powerful tools in astrophysics and cosmology. As early as 1921, the gravitational deflection of light due to the Sun was employed to act as the first test of general relativity~\cite{DED1920,Will2015}. Applications of gravitational lensing nowadays include measuring the mass of galaxies and clusters~\cite{Hoekstra2013,Brouwer2018,Bellagamba2019}, distinguishing between wormholes and black holes~\cite{Tsukamoto2012,Tsukamoto2013,gong2018}, detecting dark matter and dark energy~\cite{Vanderveld2012,cao2012,zhanghe2017,Huterer2018,SC2019}, and so on.

To our knowledge, there are several analytical methods devoted to studying gravitational lensing of light, containing the common geodesics approach~\cite{Weinberg1972}. Recently, Gibbons and Werner~\cite{GW2008} proposed an elegant geometrical method to study the weak gravitational deflection of light in a static and spherically symmetric spacetime. Namely, they applied the Gauss-Bonnet (GB) theorem to the corresponding optical geometry and obtained a beautiful expression to calculate the deflection angle. The significance of this method lies in the fact that it indicates the deflection angle can be regarded as a topological effect. This method was later extended to a stationary and axisymmetric spacetime by Werner~\cite{Werner2012}, in which the optical geometry is defined by the relevant Finsler-Randers metric and thus the author applied Naz{\i}m's method to construct an osculating Riemannian manifold where one could conveniently use the GB theorem. The geometrical method in Refs.~\cite{GW2008,Werner2012} was applied not only to black hole lensing~\cite{Jus-B0,Jus-B1,SO-B2,Arakida2018,Jus-BW,OJS2018,OSS2018,Jus-BW}, but also to the gravitational lensing caused by other objects such as wormholes~\cite{OJS2018,Jus-BW,Jus-W1,Jus-W2,Jus-W3,Jus-W4,Jus-W5,OV2018,Goulart2018,Javed2019}, cosmic strings~\cite{Jus-C1,Jus-C2,Jus-C3,Jus-C4,Jus-C5,AO2019}, global monopoles~\cite{Jus-M1}, or mass distributions of two-power-law densities~\cite{Leon2019}, in different gravitational theories. On the other hand, Ishihara \textit{et. al.}~\cite{ISOA2016,IOA2017-1,IOA2017-2,IOA2018,IOA2019} adopted the GB theorem to study the finite-distance corrections for gravitational deflection of light, where the source and observer were no longer assumed to be infinitely far from the lens.

Compared to the case of light, the gravitational deflections of massive particles also have extensive applications, such as analyzing the properties of massive neutrinos and cosmic rays ~\cite{Patla2014,LM2019,Marques2019}, and attract more and more attention of the relativity community~\cite{AR2002,AR2004,Bhadra2007,Yu2014,Liu2016,He2016,He2017a,He2017b,Pang2019,LZLH2019}. It is expected that the new geometrical method in Refs.~\cite{GW2008,Werner2012} can be applied to investigate massive particle lensing. Actually, several works on gravitational deflection of relativistic massive particles via the GB theorem have recently been proposed in the weak field limit. Crisnejo and Gallo~\cite{CG2018} utilized the GB theorem to study the gravitational deflections of both light in a plasma medium in a static and spherically symmetric spacetime and massive particles in Schwarzschild spacetime. This technique was later adopted to study the Reissner-Nordstr\"{o}m deflection of charged massive particles~\cite{CGV2019}. By viewing the propagating particles as the de Broglie wave packets~\cite{Evans2001}, Jusufi~\cite{Jus-massive1} calculated the deflection angles of massive particles by Kerr black hole and Teo wormhole respectively, based on the corresponding isotropic type metrics, the refractive index of the corresponding optical media, and the GB theorem. Jusufi's idea in Ref.~\cite{Jus-massive1} was further extended to distinguish naked singularities and Kerr-like wormholes~\cite{Jusufi2019-1}, and to study the gravitational deflection of charged particles in Kerr-Newman spacetime~\cite{Jusufi2019-2}.

In this work, the Jacobi metric method proposed by Gibbons~\cite{Gibbons2016} to utilize the GB theorem will be used to derive the gravitational deflection angles of relativistic neutral massive particles induced, respectively, by three types of static and spherically symmetric wormholes: the Janis-Newman-Winicour (JNW) wormhole, a class of zero Ricci scalar scalar-tensor wormholes and a class of charged Einstein-Maxwell-dilaton wormholes. The gravitational deflection angles up to the second post-Minkowskian order are obtained on the basis of the perturbation method and iterative technique. Our discussions are constrained in the weak-field, small-angle, and thin-lens approximation.

This paper is organized as follows. In Sec.~\ref{TF}, we shall set up the general framework for our calculations, including the static and spherically symmetric Jacobi metric, the GB theorem applied to the Jacobi geometry, and the asymptotically Euclidean case. In Sec.~\ref{App}, we derive the gravitational deflection angle of relativistic massive particles in three types of static and spherically symmetric wormhole spacetimes. Finally, we summarize our results in Sec.~\ref{CONCLU}. Throughout this paper, we use the natural units where $G = c = 1$ and the metric signature $(-,+,+,+)$. For convenience, $g_{ij}$ is used to denote Jacobi metric if no confusion is caused, and this is followed for the quantities with Jacobi metric, while the quantities associated with the background spacetime metric are added a bar above.

\section{static and spherically symmetric Jacobi metric and the Gauss-Bonnet theorem}
\label{TF}

\subsection{Static and spherically symmetric Jacobi metric } \label{orbit equation}
According to the principle of least action of Maupertuis, Gibbons \textit{et. al.}~\cite{Gibbons2016,Gibbons2017-1,Gibbons2017-2,Gibbons2019} established the Jacobi metric framework for curved spacetime. The motion of free massive particles in background spacetime can be described as a spatial geodesic in the corresponding Jacobi geometry defined by the Jacobi metric, which is similar to the case where the motion of the photon can be described as a spatial geodesic in the corresponding optical geometry. Even for charged particles~\cite{Das2017}, the Jacobi metric approach also works. For this reason, one can use the Jacobi geometry as a background space to study the deflection of particles.

For a static metric
\begin{eqnarray}
d\bar s^2=\bar g_{tt}dt^2+\bar g_{ij}dx^i dx^j,
\end{eqnarray}
the corresponding Jacobi metric reads~\cite{Gibbons2016}
\begin{eqnarray}
\label{sjmetric}
&& g_{ij}=\left(E^2+m^2 \bar g_{tt}\right) g^{\mathrm{opt}}_{ij},
\end{eqnarray}
where $E$ and $m$ are the particle energy and mass respectively, and $g^{\mathrm{opt}}_{ij}$ is the corresponding optical metric of the static metric given by~\cite{GW2008}
\begin{eqnarray}
&& g^{opt}_{ij}=-\frac{\bar g_{ij}}{\bar g_{tt}}.
\end{eqnarray}
Notice that the Jacobi metric in Eq.~\eqref{sjmetric} is actually the same with a special optical metric related with massive particles in Ref.~\cite{CG2018}.

The general form for a static and spherically symmetric metric is written as
\begin{equation}
\label{SS-metric}
d\bar s^2=-A\left(r\right)dt^2+B\left(r\right)dr^2+C\left(r\right)d\Omega^2,\\
\end{equation}
where $d\Omega^2=d\theta^2+\sin^2\theta d\varphi^2$ is the line element of the unit two-sphere. By Eq.~\eqref{sjmetric}, its corresponding Jacobi metric is
\begin{eqnarray}
\label{SS-Jacobi-1}
ds^2&=&\bigg(E^2-m^2A\bigg)\bigg[\frac{B}{A}dr^2+\frac{C}{A}d\Omega^2 \bigg].
\end{eqnarray}
Due to spherical symmetry, we study only the motion of massive particles in the equatorial plane $\theta=\pi/2$ without loss of generality. Thus, the Jacobi metric becomes
\begin{equation}
\label{SS-Jacobi-2}
ds^2=\bigg(E^2-m^2A\bigg)\bigg(\frac{B}{A}dr^2+\frac{C}{A}d\varphi^2 \bigg).\\
\end{equation}
Then, one can obtain the conserved angular momentum $J$ by axial symmetry
\begin{equation}
\label{angular momentum}
J=\left(E^2-m^2A\right)\frac{C}{A}\left(\frac{d\varphi}{ds}\right)= \mathrm{constant},
\end{equation}
together with Eq.~\eqref{angular momentum} and Eq.~\eqref{SS-Jacobi-2}, which yields
\begin{equation}
\label{radial equation}
\left(E^2-m^2A\right)^2\frac{B}{A}\left(\frac{dr}{ds}\right)^2=E^2-A\left(m^2+\frac{J^2}{C}\right).
\end{equation}
This is consistent with the standard result
\begin{equation}
m^2A B\left(\frac{dr}{d\tau}\right)^2=E^2-A\left(m^2+\frac{J^2}{C}\right),
\end{equation}
where $\tau$ is used to denote the proper time along the geodesic, and then
\begin{eqnarray}
E=m A\frac{dt}{d\tau}\label{conserved quantity 1}~,\ \ J=m C\frac{d\varphi}{d\tau},
\end{eqnarray}
with
\begin{equation}
\label{proper time}
d\tau=\frac{m A}{E^2-m^2A}ds.
\end{equation}
Introducing the inverse radial coordinate $u=1/r$, the orbit equation can be obtained from Eqs.~\eqref{angular momentum} and~\eqref{radial equation} as follows:
\begin{eqnarray}
\label{trajectory equation}
\left(\frac{du}{d\varphi}\right)^2=\frac{C^2u^4}{AB}\left[\left(\frac{\varepsilon}{h}\right)^2-A\left(\frac{1}{h^2}+\frac{1}{C}\right)\right]~,
\end{eqnarray}
where $h=J/m$ is the angular momentum per unit mass and $\varepsilon=E/m$ is the energy per unit mass. The energy and angular momentum for an asymptotic observer  at infinity are~\cite{CG2018}
\begin{eqnarray}
\label{enan}
&&E=\frac{m}{\sqrt{1-v^2}}~,~~J=\frac{m v b}{\sqrt{1-v^2}},
\end{eqnarray}
where $v$ is the particle velocity and $b$ is the impact parameter defined by
\begin{eqnarray}
\label{impact}
\frac{J}{E}=vb.
\end{eqnarray}
Via Eq.~\eqref{enan}, Jacobi metric~\eqref{SS-Jacobi-2} becomes
\begin{equation}
\label{Jacobi-metric}
ds^2=m^2\bigg(\frac{1}{1-v^2}-A\bigg)\bigg[\frac{B}{A}dr^2+\frac{C}{A}d\varphi^2 \bigg]~,
\end{equation}
and the trajectory equation~\eqref{trajectory equation} comes to
\begin{eqnarray}
\label{trajectory equation1}
\left(\frac{du}{d\varphi}\right)^2&=&\frac{C^2u^4}{AB}\bigg[\frac{1}{v^2b^2}-A\left(\frac{1-v^2}{v^2 b^2}+\frac{1}{C}\right)\bigg].
\end{eqnarray}

\subsection{The Gauss-Bonnet theorem and lens geometry}
\label{The Jacobi metric1}

\begin{figure}[t]
\centering
\includegraphics[width=6.0cm]{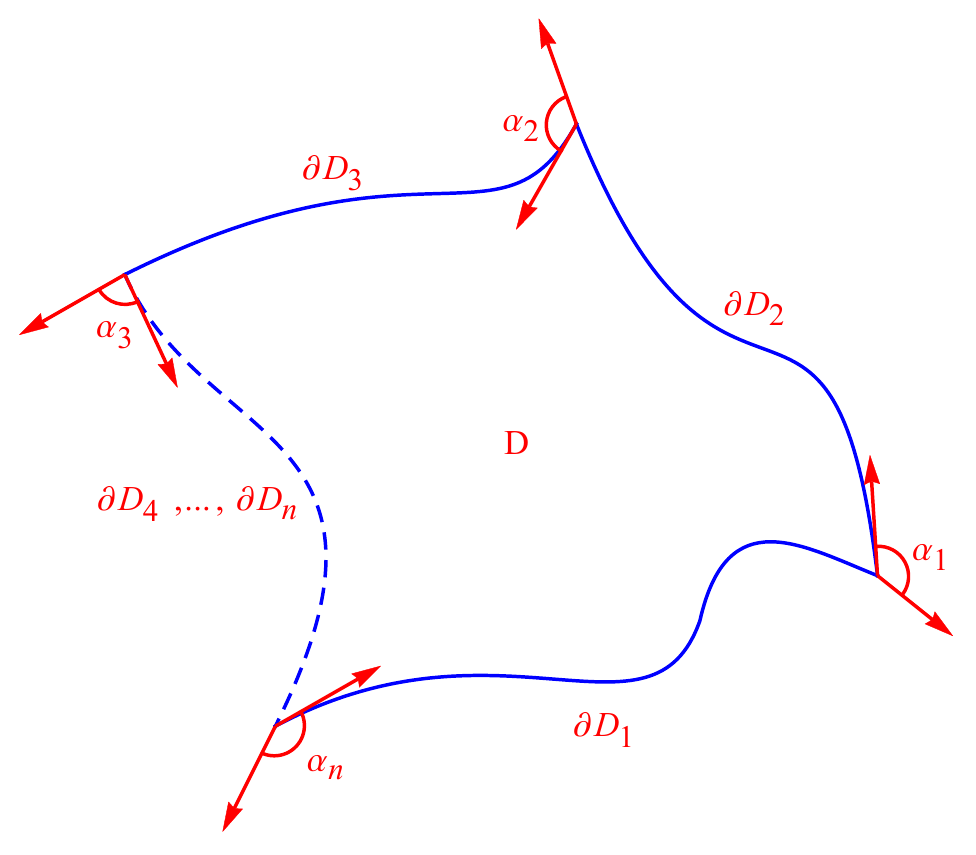}
\caption{A region $D$ with boundary $\partial D=\bigcup_i \partial D_i$. $\alpha_i$ is the exterior angle at the $i$th vertex in the positive sense.}\label{Figure1}
\end{figure}
Let $D$ be a compact oriented surface with a Riemannian metric $\hat{g}_{ij}$, Gaussian curvature $K$, and Euler characteristic $\chi(D)$. Its boundary $\partial{D}:\mathbb{R}\supset{I}\rightarrow{D}$ is a piecewise smooth curve with geodesic curvature $k_g$. The GB theorem states that~\cite{GW2008}
\begin{equation}
\label{GBT}
\iint_D{K}dS+\oint_{\partial{D}}k_g~dl+\sum_i{\alpha_i}=2\pi\chi(D),\\
\end{equation}
where $dS$ is the area element of the surface, $dl$ is the line element along the boundary, and $\alpha_i$ is the exterior angle defined for the $i$th vertex in the positive sense, as shown in Fig.~\ref{Figure1}.

\begin{figure}[t]
\centering
\includegraphics[width=7.0cm]{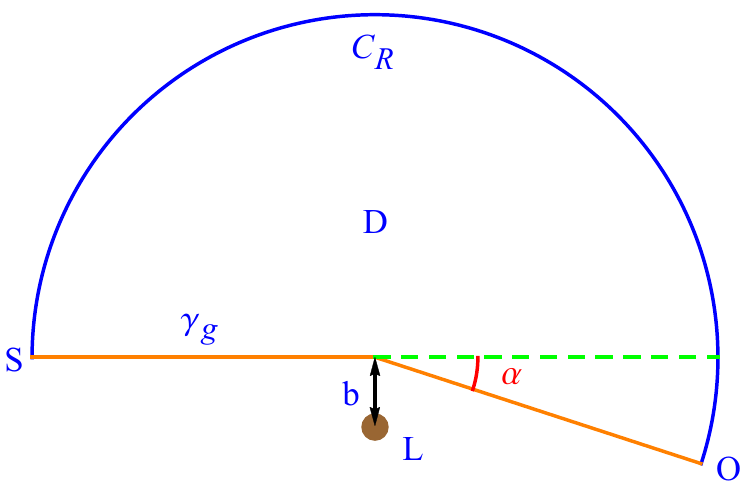}
\caption{The region $D$ belongs to the two-dimensional Jacobi space with boundary $\partial D=\gamma_g \bigcup C_R$. The particle trajectory $\gamma_g$ is a spatial geodesic and $ C_R$ is a curve defined by $r(\varphi)=R=\mathrm{constant}$. $S$, $O$, and $L$ denote the particle source, the observer, and the gravitational lens, respectively. $\alpha$ is the deflection angle and $b$ is the impact parameter. Note that $S$ and $O$ are both assumed to be at infinite distance from $L$ in our lensing setup (i.e. $R\rightarrow\infty$).}\label{Figure2}
\end{figure}

Next, the GB theorem will be employed to the Riemann-Jacobi geometry defined by~Eq.~\eqref{Jacobi-metric}. Consider a Jacobi region $D$ with boundary $\partial{D}=\gamma_g \bigcup C_R$. Here $C_R$ is a curve defined by $r(\varphi)=R=\mathrm{constant}$, which intersects the particle trajectory $\gamma_g$ at two points, the source $S$ and the observer $O$, respectively. $\gamma_g$ is a spatial geodesic leading to $k_g(\gamma_g)=0$, and one has $\chi(D)=1$ because the region $D$ does not contain the gravitational lens $L$. This paper mainly focuses on the deflection angle for the source and the observer at infinite distance from the lens. Notice that $\alpha_O+\alpha_S\longrightarrow{\pi}$ as $R\rightarrow{\infty}$, and then applying the GB theorem to region $D$ leads to
\begin{equation}
\label{general formula}
\lim_{R\rightarrow\infty}\int_{0}^{\pi+\alpha}\left({k _g\frac{ds}{d\varphi}}\right)\bigg{|}_{C_R}d\varphi=\pi-\lim_{R\rightarrow\infty}\iint_D{K}dS~,
\end{equation}
with the deflection angle $\alpha$ (which can be described by impact parameter $b$) shown in Fig.~\ref{Figure2}.

The Gaussian curvature with respect to Jacobi metric $g_{ij}$ can be calculated by~\cite{Werner2012}
\begin{eqnarray}
\label{Gauss-K}
&&K=\frac{1}{\sqrt{\det g}}\left[\frac{\partial}{\partial{\varphi}}\left(\frac{\sqrt{\det g}}{g_{rr}}{{\Gamma}^\varphi_{rr}}\right)-\frac{\partial}{\partial{r}}\left(\frac{\sqrt{\det g}}{g_{rr}}{{\Gamma}^\varphi_{r\varphi}}\right)\right],\nonumber\\
\end{eqnarray}
where $\det g$ denotes the determinant of Jacobi metric and ${\Gamma}^i_{j k}$ is the Christoffel symbol.
From Jacobi metric in Eq.~\eqref{Jacobi-metric}, one has
\begin{eqnarray}
\label{dds}
\frac{ds}{d\varphi}\bigg{|}_{C_R}=\left[m^2\bigg(\frac{1}{1-v^2}- A(R)\bigg)\frac{C(R)}{A(R)}\right]^{1/2}.
\end{eqnarray}
Furthermore, one can choose the velocity along the curve $C_R$ as $\dot{C}_{R}^i=\left(0,d\varphi (R)/ds\right)$, which satisfies unit speed condition $g_{ij}\dot{C}_{R}^{i}\dot{C}_{R}^{j}=1$.
This condition yields
\begin{eqnarray}
&&(\nabla_{\dot{C}_R}{\dot{C}_R})^{r}=\Gamma_{\varphi\varphi}^{r}(R)(\dot{C}_{R}^{\varphi})^2~,~~~(\nabla_{\dot{C}_R}{\dot{C}_R})^{\varphi}=0,
\end{eqnarray}
and now, the geodesic curvature of $C_R$ can be expressed as follows
\begin{eqnarray}
\label{geode-cur}
k_g(C_R)=\mid\nabla_{\dot{C}_R}{\dot{C}_R}\mid=\sqrt{g_{rr}\left(\Gamma_{\varphi\varphi}^{r}\right)^2}\left(\frac{d\varphi}{ds}\right)^2\bigg{|}_{C_R}.
\end{eqnarray}
Together with Eqs.~\eqref{dds} and~\eqref{geode-cur}, one can obtain
\begin{eqnarray}
&&\left({k_g\frac{ds}{d\varphi}}\right)\bigg{|}_{C_R}=\sqrt{\frac{B(R)}{C(R)}\left(\Gamma_{\varphi\varphi}^{r}(R)\right)^2}~. \label{geodesic}
\end{eqnarray}

\subsection{Asymptotically Euclidean space}
At this point, a special case can be considered
\begin{eqnarray}
\label{condition}
\lim_{R\rightarrow\infty}\left({k_g\frac{ds}{d\varphi}}\right)\bigg{|}_{C_R}=1,
\end{eqnarray}
which means that the two-dimensional Jacobi geometry in Eq.~\eqref{Jacobi-metric} is asymptotically Euclidean. Then, Eq.~\eqref{general formula} leads to
\begin{eqnarray}
\label{GBT-K}
\alpha=-\lim_{R\rightarrow\infty}\iint_D{K}dS.
\end{eqnarray}
This expression is the same to the result obtained by applying the GB theorem to the optical metric~\cite{GW2008,Werner2012,CG2018,Jus-massive1}. From Eq.~\eqref{GBT-K}, the deflection angle of massive particles can be obtained by integrating the intrinsic curvature of space, and the integral region is an infinite Jacobi domain outside the particle ray relative to the lens. Therefore, the deflection angle can be regarded as a global topological effect~\cite{Werner2012,Jus-massive1}.

In this paper, we mainly focus on the weak deflection limit and the deflection angle up to second order in lens parameter $\varepsilon$ is calculated. For this purpose, the iterative method is reasonable to carry out, and we first consider the first-order approximation. Notice that the Gaussian curvature contains at least the first-order term, and thus it is sufficient to use only the zero-order particle trajectory $r=b/\sin(\varphi),~0\leq\varphi\leq{\pi}$. As a result, Eq.~\eqref{GBT-K} can be expressed as
\begin{eqnarray}
\label{first-order}
\alpha_1\approx-\int_{0}^{\pi}\int_{b/\sin{\varphi}}^{\infty}{K\sqrt{\det g}}~dr d\varphi.
\end{eqnarray}
Moreover, the second-order deflection angle can be expressed as
\begin{eqnarray}
\label{GBT-K2}
\alpha\approx-\int_{0}^{\pi+\alpha_1}\int_{b/\sin{\varphi}+r_1(\varphi)\varepsilon}^{\infty}{K\sqrt{\det g}}~drd\varphi,
\end{eqnarray}
where the first-order trajectory $r=b/\sin(\varphi)+r_1(\varphi)\varepsilon,~0\leq\varphi\leq{\pi+\alpha_1}$.

In short, we can use Eq.~\eqref{GBT-K2} to calculate the second-order deflection angle in asymptotically Euclidean space. However, Eq.~\eqref{general formula} is required if the Jacobi metric is not asymptotically Euclidean. Therefore, before calculating the deflection angle, one should first check whether the Eq.~\eqref{condition} is satisfied for a specific spacetime.

\section{deflection angle of massive particles by wormholes}
\label{App}
A wormhole is a speculative structure connecting far-separated spacetime points, predicted by a special solution of the Einstein field equations in general relativity and modified theories of gravity. Though there is no direct experimental evidence that wormholes exist, the wormholes may be created in the early Universe, and the original wormholes may have survived to these days~\cite{HM1981,DA1991,LZ2014}. Since wormholes can explain typical phenomena usually attributed to black holes, the objects generally considered to be black holes in the center of galaxies may be wormholes created in the early Universe~\cite{LZ2014}. For this reason, some authors have proposed different methods to probe wormholes~\cite{TN2016,Sabin2017,DD2019}. In addition, how to distinguish black hole, wormhole and naked singularity was studied in Refs.~\cite{Tsukamoto2012,Tsukamoto2013,gong2018,Jusufi2019-1}. In these studies, the gravitational lensing as a basic tool was considered. In the following, we will consider the gravitational deflection of relativistic massive particles by three types of wormholes, and this may be helpful to the studies on the wormholes.

\subsection{The JNW wormhole} \label{GBTDA}
The famous JNW wormhole, a class of static and spherically symmetric exact solutions for the Einstein minimally coupled scalar theory, is given by~\cite{Nandi2006,Dey2008}
\begin{eqnarray}
A\left(r\right)&=&B\left(r\right)^{-1}=\left(1-\frac{2M_J}{r}\right)^\gamma,\nonumber\\
C\left(r\right)&=&r^2\left(1-\frac{2M_J}{r}\right)^{1-\gamma}, \label{JNW-ABC}
\end{eqnarray}
with $\gamma=M_\mathrm{ADM}/M_J$. Here $M_\mathrm{ADM}$ is the ADM mass related to the asymptotic scalar charge $q$ by $M_\mathrm{ADM}^2=M_J^2-kq^2/2$, where $k>0$ is the matter-scalar field coupling constant. This solution corresponds to the naked singularity for $\gamma<1$ with real scalar charge, to the Schwarzschild black hole for $\gamma=1$ with zero scalar charge, and to the wormhole for $\gamma>1$ with complex scalar charge~\cite{Nandi2006}. Recently, Formiga and Almeida have shown that this wormhole ($\gamma>1$) cannot be traversed by humans, but it can be traversed by particles and objects that last long enough~\cite{JT2014}.

Substituting~\eqref{JNW-ABC} into~\eqref{Jacobi-metric}, one can find the JNW-Jacobi metric induced in the equatorial plane as follows:
\begin{eqnarray}
\label{JNW-Jacobi}
ds^2&=&m^2\left[\frac{1}{1-v^2}-\left(1-\frac{2M_J}{r}\right)^\gamma\right]\nonumber\\
&& \times \left[\frac{dr^2}{\left(1-\frac{2M_J}{r}\right)^{2\gamma}}+\frac{r^2d\varphi^2}{\left(1-\frac{2M_J}{r}\right)^{2\gamma-1}} \right],
\end{eqnarray}
with determinant
\begin{eqnarray}
\label{JNW-det}
\det{g}=\frac{m^4 r^2 \left[1-\left(1-v^2\right)\left(1-\frac{2M_J}{r}\right)^{\gamma}\right]^2}{(1-v^2)^2(1-\frac{2M_J}{r})^{4\gamma-1}}.
\end{eqnarray}
The Gaussian curvature up to second order in $M_J$ can be obtained by Eq.~\eqref{Gauss-K} as follows:
\begin{eqnarray}
\label{JNW-2K}
\nn K&=&-\frac{1-v^2}{m^2 v^2 r^2}\Bigg \{\frac{\left(1+v^2\right)\gamma M_J}{v^2r}-\bigg[1-\frac{3\left(1+v^2\right)\gamma}{v^2}\\
&&+\frac{2\left(3+v^4\right)\gamma^2}{v^4}\bigg]\frac{M_J^2}{r^2}\Bigg \}+\mathcal{O}\left(M_J^3\right).
\end{eqnarray}
Next, we will check if the JNW-Jacobi geometry is asymptotically Euclidean. Considering Eq.~\eqref{geodesic}, by some algebra, one has
\begin{eqnarray}
\nn\left({k_g\frac{ds}{d\varphi}}\right)\bigg{|}_{C_R}&=&\sqrt{\frac{B(R)}{C(R)}\left(\Gamma_{\varphi\varphi}^{r}(R)\right)^2}\\
\nn&=&\frac{1}{\sqrt{\left(1-\frac{2M_J}{R}\right)}R}\bigg[R-\left(1+\gamma\right)M_J\\
&&-\frac{\gamma M_J}{1-\left(1-v^2\right)\left(1-\frac{2M_J}{R}\right)^\gamma}\bigg],
\end{eqnarray}
which leads to
\begin{eqnarray}
\lim_{R\rightarrow\infty}\left({k_g\frac{ds}{d\varphi}}\right)\bigg{|}_{C_R}=1~,
\end{eqnarray}
and this means that JNW-Jacobi metric Eq.~\eqref{JNW-Jacobi} is asymptotically Euclidean and Eq.~\eqref{GBT-K2} is efficient to calculate the deflection angle for JNW spacetime.

By Eq.~\eqref{first-order}, the first-order deflection angle can be obtained as follows:
\begin{eqnarray}
\label{1jnw}
\alpha_1&\approx&-\int_{0}^{\pi}\int_{b/\sin{\varphi}}^{\infty}{K\sqrt{\det g}}~dr d\varphi\nn \\
&\approx&\int_{0}^{\pi}\int_{b/\sin{\varphi}}^{\infty}{\frac{\left(1+v^2\right)\gamma M_J}{v^2 r^2}}~dr d\varphi\nn \\
&=&\frac{2\left(1+v^2\right)\gamma M_J}{bv^2}.
\end{eqnarray}
On the other hand, the first-order trajectory of massive particles obtained by perturbation method is discussed in Appendix~\ref{PTT}. Then, based on the second-order Gaussian curvature in Eq.~\eqref{JNW-2K}, the first-order particle ray in Eq.~\eqref{JNW-orbit} and the deflection angle $\alpha_1$ in Eq.~\eqref{1jnw}, the second-order deflection angle can be obtained according to Eq.~\eqref{GBT-K2}
\begin{eqnarray}
\nn&&\alpha=\frac{2\left(1+v^2\right)\gamma M_J}{bv^2}+\left(1+\frac{3}{v^2}-\frac{1}{4\gamma^2}\right)\frac{\pi \gamma^2 M_J^2 }{b^2}  \\
&&~~~~~~~+\mathcal{O}\left(M_J^3\right).  \label{JNW-2mdef}
\end{eqnarray}
Now, two limiting cases for Eq.~\eqref{JNW-2mdef} can be considered. First, for $v=1$, the second-order deflection angle for light can be recovered,
\begin{eqnarray}
\label{JNW-light}
\alpha=\frac{4\gamma M_J}{b}+\frac{\left(-1+16\gamma^2\right)\pi M_J^2}{4b^2}+\mathcal{O}\left(M_J^3\right),
\end{eqnarray}
which is in agreement with the result in Refs.~\cite{Virbhadra2008,Gyulchev2008}. Here, one should  notice that only the first-order term is consistent with the results by Jusufi~\cite{Jus-W1} using the GB theorem, and this difference comes from the fact that the straight line approximation $r\left(\varphi\right)=b/\sin\varphi$ was used in Ref.~\cite{Jus-W1}, where the first-order perturbation term of gravity was ignored. In the latest work~\cite{Jusufi2019-1}, Jusufi~\textit{et al.} used the geodesics approach to obtain the consistent result with Eq.~\eqref{JNW-2mdef}.

Second, for $\gamma=1$ and $M_J=M_S$, the Schwarzschild spacetime can be recovered, and the second-order deflection angle becomes
\begin{eqnarray}
\label{JNW-S}
\alpha_S=\frac{2\left(1+v^2\right) M_S}{bv^2}+\frac{3\left(4+v^2\right)\pi M_S^2}{4b^2v^2}+\mathcal{O}\left(M_S^3\right),
\end{eqnarray}
where $M_S$ is the mass of the Schwarzschild black hole.
Equation~\eqref{JNW-2mdef} shows that the deflection angle increases as $\gamma$ increases. Therefore, it is obvious that $\alpha_\mathrm{worm} > \alpha_S  > \alpha_\mathrm{sing}$ for $M_J=M_S$ and the same $v$ and $b$, where $\alpha_\mathrm{worm}$ and $\alpha_\mathrm{sing}$ are the second-order deflection angles for the wormhole and naked singularity, respectively. This difference may be used to distinguish the black hole from wormhole or naked singularity. For the study on the discrimination between black hole, wormhole and naked singularity by the gravitational lensing, we refer the reader to Refs.~\cite{Tsukamoto2012,Tsukamoto2013,gong2018,Jusufi2019-1}. Moreover, for massive particles in Schwarzschild spacetime, there are two different results, one obtained by Accioly and Ragusa~\cite{AR2002} and the other by Bhadra \textit{et al.}~\cite{Bhadra2007}. Our expression~\eqref{JNW-S}, with other work by He and Lin~\cite{He2016} using the post-Minkowskian iterative method and by Crisnejo and Gallo~\cite{CG2018} using GB theorem, is in agreement with the result by Accioly and Ragusa~\cite{AR2002}.

\subsection{ A class of $\bar{R}=0$ scalar-tensor wormholes}
In this subsection, we consider a class of wormholes with $\bar{R}=0$ in the context of the scalar-tensor theory of gravity, where $\bar{R}$ is the Ricci scalar. Shaikh and Kar~\cite{Shaikh2016} first obtained these solutions and subsequently studied the deflection angle of light~\cite{Shaikh2017}. For Shaikh-Kar wormhole, one has~\cite{Shaikh2016,Shaikh2017}:
\begin{eqnarray}
\label{0R-ABC}
\nn A\left(r\right)&=&\frac{\left[1+\frac{\beta}{M^2}\frac{ M }{r}+\eta\sqrt{1-\frac{2 M }{r}-\frac{\beta}{M^2}\frac{ M^2 }{r^2}}\right]^2}{(1+\eta)2},\\
\nn B\left(r\right)&=&\left(1-\frac{2 M }{r}- \frac{\beta}{M^2}\frac{ M^2 }{r^2}\right)^{-1},\\
C\left(r\right)&=&r^2,
\end{eqnarray}
where $\beta/M^2 =(1-\mu^2)/\mu^2$ with $M$ being the ADM mass, and $\mu$ and $\eta$ are two constants. These solutions correspond to the naked singularities when $n<-1$ and to the traversable wormholes when $n>-1$~\cite{Shaikh2016,Shaikh2017}.
Now, Shaikh-Kar-Jacobi metric induced in the equatorial plane can be obtained by Eq.~\eqref{Jacobi-metric} as follows:
\begin{equation}
\label{JST}
ds^2=m^2\bigg[\frac{1}{1-v^2}-A(r)\bigg]\frac{B(r)dr^2+r^2d\varphi^2 }{A\left(r\right)}.
\end{equation}
One can obtain the corresponding Gaussian curvature by Eq.~\eqref{Gauss-K} and the result to second order is
\begin{eqnarray}
\label{0R-2K}
K&=&-\frac{1-v^2}{m^2 v^2 r^2}\bigg \{\left[1+v^2-\frac{1}{(1+\eta)\mu^2}\right]\frac{M}{rv^2}\nonumber\\
&&\nn+\bigg[1-v^2-\frac{6}{v^2}+\frac{2+v^2}{\mu^2}-\frac{6\left(1-v^2\right)}{v^2\left(1+\eta\right)^2\mu^4}\\
&&+\frac{3\left(4-3v^2\right)}{v^2\left(1+\eta\right)\mu^2}\bigg]\frac{M^2}{r^2v^2}\bigg \}+\mathcal{O}(M^3).
\end{eqnarray}

Similar to the last subsection, one can easily verify that Eq.~\eqref{condition} holds here, which implies Shaikh-Kar-Jacobi geometry is asymptotically Euclidean and thus the deflection angle of massive particles can be obtained by Eq.~\eqref{GBT-K2}. In addition, the first-order particle trajectory is given in Appendix~\ref{PTT}. Finally, after calculating the first-order deflection angle $\alpha_1$ by Eq.~\eqref{first-order}, considering the Gauss curvature~\eqref{0R-2K} and particle ray~\eqref{OR-orbit}, the deflection angle up to second order can be obtained as follows:
\begin{eqnarray}
\label{0R-2def}
\nn \alpha &=&\left[1+v^2-\frac{1}{(1+\eta)\mu^2}\right]\frac{2M}{b v^2}\\
&&\nn +\bigg[5+v^2+\frac{2+v^2}{2\mu^2}-\frac{9}{(1+\eta)\mu^2}\\
&&+\frac{3}{(1+\eta)^2\mu^4}\bigg]\frac{\pi M^2}{2b^2v^2}+\mathcal{O}(M^3).
\end{eqnarray}
This result shows that the naked singularities $(\eta<-1)$ add a positive term to the deflection angle and the deflection angle is always positive, whereas the wormhole solutions $(\eta>-1)$ add a negative term to the deflection angle, at each order.

For the photon $v=1$, expression~\eqref{0R-2def} reduces to
\begin{eqnarray}
\nn \alpha &=&\left[1-\frac{1}{2\mu^2(1+\eta)}\right]\frac{4M}{b}\\
&&\nn+\bigg[1+\frac{1}{4\mu^2}-\frac{3}{2\mu^2(1+\eta)}\\
&&+\frac{1}{2\mu^4(1+\eta)^2}\bigg]\frac{3\pi M^2}{b^2}+\mathcal{O}(M^3),
\end{eqnarray}
which is consistent with the results in Ref.~\cite{Shaikh2017}. In the limit $\eta\rightarrow{\infty}$ ,$M=M_\mathrm{RN}$ and $\mu^2\rightarrow{\frac{1}{1-q^2/M_\mathrm{RN}^2}}$, the result for the Reissner-Nordstr\"{o}m black hole can be recovered
\begin{eqnarray}
\nn {\alpha_\mathrm{RN}}&=&\frac{2\left(1+v^2\right)M_\mathrm{RN}}{bv^2}+\frac{3\left(4+v^2\right)\pi M_\mathrm{RN}^2}{4 b^2v^2}\\
&&-\frac{\left(2+v^2\right)\pi q^2}{4b^2 v^2}+\mathcal{O}(M_\mathrm{RN}^3, q^4),
\end{eqnarray}
where $M_\mathrm{RN}$ and $q$ are the mass and electrical charge of the Reissner-Nordstr\"{o}m black hole, respectively. This expression is in agreement with the results in Refs.~\cite{Pang2019,He2017b}. One can also obtain the result in Eq.~\eqref{JNW-S} for Schwarzschild spacetime as $\eta\rightarrow{\infty}$ , $M=M_S$ and $\mu^2\rightarrow{1}$.

\subsection{A class of charged Einstein-Maxwell-dilaton wormholes}
In this subsection, we consider a class of charged wormholes which arise as solutions in the Einstein-Maxwell-dilaton theory obtained by Goulart~\cite{Goulart2017} as follows:
\begin{eqnarray}
\label{EMD-ABC}
\nn A\left(r\right)&=&\frac{r^2}{r^2+2PQ},\\
\nn B\left(r\right)&=&\frac{r^2+2PQ}{r^2+\Sigma^2+2PQ},\\
C\left(r\right)&=&r^2+2PQ,
\end{eqnarray}
where $P$ is the magnetic charge, $Q$ is the electric charge and $\Sigma$ is the dilaton charge. There are two interesting facts about Goulart wormhole. First, it is shown that the massless solution seems physically acceptable. Second, it is also shown that a wormhole satisfying the null energy condition in the classical theory can be constructed. Furthermore, it is meaningful to study the traversability of this wormhole~\cite{Goulart2017}.

For Goulart wormhole, the leading-order term of deflection angle for light was obtained by Jusufi~\cite{Jus-W2} with the Gibbons-Werner method and by Lukmanova \textit{et al.}~\cite{Lukmanova2018} with the parametric post-Newtonian method as follows:
\begin{eqnarray}
\label{JL}
\alpha=\frac{3\pi P Q}{2b^2}-\frac{\pi \Sigma^2}{4b^2}+\mathcal{O}\big(P^2,Q^2,\Sigma^4\big).
\end{eqnarray}
Obviously, the electric charge $Q$ and the magnetic charge $P$ contribute to the deflection angle, whereas the dilaton charge $\Sigma$~ decreases it~\cite{Jus-W2}. In order to see the effect of the dilaton in more details on deflection angle of massive particles, it is expected to obtain the result of orders $PQ,\Sigma^2,P^2Q^2,PQ\Sigma^2$. Thus, the term of $\Sigma^2$ cannot be ignored for particle trajectory and this is dealt with in Appendix~\ref{PTT}.

Substitution Eq.~\eqref{EMD-ABC} into Eq.~\eqref{Jacobi-metric}, one can deduce the Goulart-Jacobi metric in the equatorial plane as follows:
{\begin{eqnarray}
\label{EMD-Jacobi}
ds^2&=&m^2\left(r^2+2PQ\right)^2\bigg(\frac{1}{1-v^2}-\frac{ r^2}{r^2+2PQ}\bigg)\nn\\
&& \times\left[\frac{dr^2}{r^2(r^2+\Sigma^2+2PQ)}+\frac{1}{r^2}d\varphi^2\right],
\end{eqnarray}}
with the Gaussian curvature
{\begin{eqnarray}
\label{EMD-K}
 K&=&\frac{1-v^2}{m^2 r^4 v^2}\bigg[\Sigma^2-\frac{2\left(2+v^2 \right)PQ}{v^2}\nn\\
&&-\frac{8\left(1+v^2 \right)PQ \Sigma^2}{r^2v^2}+\frac{8\left(3+v^4\right)P^2Q^2}{r^2v^4}\bigg]\nn\\
&&+\mathcal{O}\big(P^3,Q^3,\Sigma^4\big).
\end{eqnarray}}

For Goulart-Jacobi geometry, one can easily check that Eq.~\eqref{condition} also holds, and thus the deflection angle can be derived by Eq.~\eqref{GBT-K2}. We calculate the leading order of deflection angle of massive particles $\alpha_1$ by Eq.~\eqref{first-order} and then consider the Gaussian curvature~\eqref{EMD-K} and particle ray~\eqref{EMD-r}. Finally, the deflection angle can be obtained by Eq.~\eqref{GBT-K2} as follows:
\begin{eqnarray}
\label{EMD-R}
\alpha&=&\frac{\pi(2+v^2) P Q}{2b^2 v^2}-\frac{\pi \Sigma^2}{4b^2}-\frac{3\pi(4+v^2)P Q \Sigma^2}{16b^4v^2}\nn\\
&&+\frac{3\pi(8+24v^2+3v^4)P^2Q^2}{16b^4v^4}+\mathcal{O}\big(P^3,Q^3,\Sigma^4\big).
\end{eqnarray}
From Eq.~\eqref{EMD-R} one can observe that the terms containing $\Sigma^2$ are all negative no matter they contain $P$ and $Q$ or not, while the terms without $\Sigma^2$ are always positive.

Considering the deflection angle for light, Eq.~\eqref{EMD-R} leads to
\begin{eqnarray}
\alpha&=&\frac{3\pi P Q}{2b^2}-\frac{\pi \Sigma^2}{4b^2}-\frac{15\pi P Q \Sigma^2}{16b^4}\nn\\
&&+\frac{105\pi P^2Q^2}{16b^4}+\mathcal{O}\big(P^3,Q^3,\Sigma^4\big),
\end{eqnarray}
and the leading-order terms are consistent with Eq.~\eqref{JL}.

\section{conclusion} \label{CONCLU}
In this work, we study the gravitational deflection of relativistic neutral massive particles in static and spherically symmetric wormhole spacetimes. First, we derive the static and spherically symmetric Jacobi metric and the corresponding orbit equation in the equatorial plane. Second, following Gibbons and Werner~\cite{GW2008}, we use the GB theorem to the Jacobi geometry and obtain the expression to calculate the deflection angle from Gaussian curvature. In particular, we focus on the asymptotically Euclidean Jacobi space and show that the deflection angle can be viewed as a topological effect. Finally, we study in detail the gravitational deflection in three types of wormholes: JNW wormhole, a class of $\bar{R}=0$ scalar-tensor wormholes, and a class of charged Einstein-Maxwell-dilaton wormholes.

The deflection angles up to the second post-Minkowskian order are obtained in Eqs.~\eqref{JNW-2mdef},~\eqref{0R-2def} and~\eqref{EMD-R}, where, the perturbation method is applied to obtain the first-order particle trajectory and then iterative technique is available to obtain these results. According to Eq.~\eqref{GBT-K2}, first-order deflection angle is necessary to calculate the second-order deflection angle in the iterative procedure. Thus, to calculate a higher-order angle, one first needs to obtain low-order results, which makes the calculation cumbersome. How to avoid this iteration is an interesting question and we will leave this as our future work. In addition, the influence of the spacetime parameters for different gravitational theories on the results are analyzed. In Refs.~\cite{Jus-W1,Jus-W2,Jusufi2019-1}, the authors show that the method using the GB theorem is consistent with standard geodesics method in the first-order terms. Compared with the results in these previous literatures, our work shows that the equivalence between the geometric method and other methods also holds in the second-order terms. Furthermore, it would be interesting to see if this geometric method can calculate the exact deflection angle.

In summary, this work contributes to the applications of Jacobi metric approach. In the future work, we hope that the same method can be employed to investigate the stationary spacetime and the finite-distance corrections, as well as the deflection of the charged particles.

\acknowledgements
This work was supported by the National Natural Science Foundation of China under Grants No.~11405136, No.~11847307 and No.~11947018, the Fundamental Research Funds for the Central Universities under Grant No.~2682019LK11, and the Research Foundation of Education Department of Hunan Province under Grant No.~18C0427.

\appendix

\section{The particle trajectories in wormhole spacetimes}\label{PTT}
Here, we mainly focus on gravitational deflection in the weak limit, and then the particle trajectory equation~\eqref{trajectory equation1} can be solved with perturbation method. Specifically, it is assumed that the solution can be expressed in powers of $\varepsilon$~\cite{CG2018,Arakida2012},
\begin{eqnarray}
\label{perturbation}
u\left(\varphi\right)=\frac{1}{b}\left[\sin{\varphi}+u_{1}\left(\varphi\right)\varepsilon+u_{2}\left(\varphi\right)\varepsilon^2+...\right].
\end{eqnarray}
Furthermore, the condition that $u\left(\varphi\right)=\frac{1}{r\left(\varphi\right)}$ takes a maximum value at $\varphi=\pi/2$ could be used.
\subsection{The JNW wormhole}
Substituting~\eqref{JNW-ABC} into~\eqref{trajectory equation1}, the orbit differential equation of massive particle in JNW equatorial plane is given by
\begin{eqnarray}
 \label{JNW-trajec}
\left(\frac{du}{d\varphi}\right)^2&=&\bigg(1-2M_J ~u\bigg)^{2\left(1-\gamma \right)} \bigg[\frac{1}{v^2b^2}-(1-2M_J~ u)^\gamma\nn\\
&&\times \left(\frac{1-v^2}{v^2 b^2}+\frac{u^2}{(1-2M_J~ u)^{1-\gamma}}\right)\bigg].
\end{eqnarray}
By letting $\varepsilon=M_J$ and substituting~\eqref{perturbation} into~\eqref{JNW-trajec}, consider the first-order term, and one can obtain a ordinary differential equation for ${u}_1\left(\varphi\right)$ as follows
\begin{eqnarray}
0&=&\dot{u}_1\left(\varphi\right)+\tan\varphi~u_1\left(\varphi\right)\nonumber\\
&&-\frac{\left(\gamma-2v^2+\gamma v^2+v^2\sin^2\varphi\right)\tan\varphi }{b v^2}=0,
\end{eqnarray}
where a dot denotes the derivative with respect to $\varphi$.
With the mentioned condition and solving the above equation, one can obtain
\begin{eqnarray}
&&u_1\left(\varphi\right)=\frac{(1+\frac{1}{v^2})\gamma-\sin^2\varphi}{b}.
\end{eqnarray}
This leads to
\begin{eqnarray}
\label{JNW-orbit}
&&r\left(\varphi\right)=\frac{b}{\sin \varphi}+\left[1-\frac{\left(1+v^2\right)\gamma}{v^2 \sin^2\varphi}\right]M_J+\mathcal{O}\left(M_J^2\right).
\end{eqnarray}
\subsection{A class of $\bar{R}=0$ scalar-tensor wormholes}
Substituting~\eqref{0R-ABC} into~\eqref{trajectory equation1}, the orbit differential equation of massive particle in Shaikh-Kar spacetime is given by
\begin{eqnarray}
\label{0R-trajec-equa}
\nn \left(\frac{du}{d\varphi}\right)^2&=&\frac{\Xi\left(u\right)}{b^2v^2}\Bigg[-1+v^2-u^2b^2v^2\\
&&+\frac{\left(1+\eta \right)^2}{\left( 1+\frac{\left( 1-\mu ^2 \right) Mu}{\mu ^2}+\eta \sqrt{\Xi\left(u\right)} \right)}\Bigg],
\end{eqnarray}
with $\Xi(u)=1-2Mu-\frac{\left( 1-\mu ^2 \right) M^2u^2}{\mu ^2} $.
Letting $\varepsilon=M$ and substituting~\eqref{perturbation} into~\eqref{0R-trajec-equa}, only keeping the first-order term, we can obtain the following differential equation:
\begin{eqnarray}
0&=&\dot{u}_1\left(\varphi\right)+\tan\varphi~u_1\left(\varphi\right)\nonumber\\
&&+\left[-1+v^2\cos^2\varphi+\frac{1}{(1+\eta)\mu^2}\right]\frac{\tan\varphi}{b v^2}.
\end{eqnarray}
It is easy to get the solution of this equation as
\begin{eqnarray}
u_1\left(\varphi\right)=\frac{1}{b v^2}\left[1+v^2\cos^2\varphi-\frac{1}{(1+\eta)\mu^2}\right].
\end{eqnarray}
Thus, the trajectory of massive particles is
\begin{eqnarray}
\label{OR-orbit}
r\left(\varphi\right)&=&\frac{b}{\sin \varphi}-\left[1+v^2\cos^2\varphi-\frac{1}{(1+\eta)\mu^2}\right]\nn \\
&&\times\frac{M}{v^2\sin^2\varphi}+\mathcal{O}\left(M^2\right).
\end{eqnarray}

\subsection{A class of charged Einstein-Maxwell-dilaton wormholes}
For Goulart wormhole, substituting~\eqref{EMD-ABC} into~\eqref{trajectory equation1}, the corresponding particle trajectory equation can be obtained
\begin{eqnarray}
\label{EMD-orbit}
\left(\frac{du}{d\varphi}\right)^2&=&\left(\frac{1}{b^2}-u^2\right)\left(1+u^2\Sigma^2\right)\nn \\
&&-\frac{2u^2\left(-1-2v^2+b^2u^2v^2\right)P Q}{b^2 v^2}\nn\\
&&+\frac{4u^4\left(2+v^2\right)P^2Q^2}{b^2v^2}+\frac{2u^4(1+v^2)PQ\Sigma^2}{b^2v^2}\nn  \\
&&+\frac{4u^6P^2Q^2\Sigma^2}{b^2v^2}+\frac{8u^6P^3Q^3}{b^2v^2}.
\end{eqnarray}

Now, $u=u\left(\varphi\right)$ can be expanded to
\begin{eqnarray}
u\left(\varphi\right)&=&\frac{1}{b}\left[\sin\varphi+u_1\left(\varphi\right)PQ+u_2\left(\varphi\right)\Sigma^2\right]\nn \\
&&+\mathcal{O}\left(P^2,Q^2,\Sigma^4\right).
\end{eqnarray}
By substituting this equation into Eq.~\eqref{EMD-orbit}, one can obtain the following equations:
\begin{eqnarray}
0&=&\dot{u}_1\left(\varphi\right)+\tan\varphi~u_1\left(\varphi\right)\nn\\
&&+\frac{\left(-1-2v^2+v^2\sin^2\varphi\right) \sin\varphi \tan\varphi}{b^2 v^2},\nn \\
0&=&\dot{u}_2\left(\varphi\right)+\tan\varphi~u_2\left(\varphi\right)-\frac{\cos \varphi \sin^2 \varphi}{2b^2}.
\end{eqnarray}
Solving the above equations, one can come to
\begin{eqnarray}
u_1\left(\varphi\right)&=&\frac{\cos\varphi}{4b^2 v^2}\bigg[(2+v^2)(\pi-2\varphi)\nn \\
&& -v^2\sin\left(2\varphi\right)+4(1+v^2)\tan\varphi\bigg],\nn\\
u_2\left(\varphi\right)&=&-\frac{\left[\pi-2\varphi+\sin\left(2\varphi\right)\right]\cos\varphi}{8b^2}.
\end{eqnarray}
This leads to the following relation
\begin{eqnarray}
\label{EMD-r}
 r\left(\varphi\right)&=&\frac{b}{\sin\varphi}+\frac{\left[\pi-2\varphi+\sin\left(2\varphi\right)\right]\cos\varphi \Sigma^2}{8b\sin^2\varphi}\nn\\
&&-\frac{\cos\varphi PQ}{4b v^2 \sin^2 \varphi}\bigg[\left(2+v^2\right)\left(\pi-2\varphi\right)\nn\\
&&-v^2\sin\left(2\varphi\right)+4\left(1+v^2\right)\tan\varphi\bigg]\nn\\
&&+\mathcal{O}\left(P^2,Q^2,\Sigma^4\right).
\end{eqnarray}

\end{document}